\documentclass[aps,pra,twocolumn,floats]{revtex4-2} %showpacs
\usepackage{amsmath}
\usepackage{graphicx}
\usepackage{dcolumn}
\usepackage{bm}
\usepackage{amssymb}
\usepackage{latexsym}
\usepackage{color}
\usepackage{subfigure}
\usepackage{ulem}

\newcommand{\veps}{\varepsilon}

\begin{document}

\title{Negative Refraction in isotropic achiral and chiral materials}

\author{Y. B. Band$^1$, Igor Kuzmenko$^{1}$, Marek Trippenbach$^2$}

\address{$^1$ Department of Chemistry, Department of Physics,
Department of Electro-Optics, and the Ilse Katz Center for Nano-Science,
Ben-Gurion University, Beer-Sheva 84105, Israel \\
$^2$ Faculty of Physics, University of Warsaw, ul. Pasteura 5, 02-093 Warszawa, Poland
}

\begin{abstract}
We show that negative refraction in materials can occur at frequencies $\omega$ where the real parts of the permittivity $\veps(\omega)$ and the permeability $\mu(\omega)$ have different sign, and that light with such frequencies can propagate just as well as light with frequencies where they have equal sign.  Therefore, for negative refraction one does not need to be in the ``double-negative'' regime.  We consider negative refractive index achiral materials using the Drude-Lorentz model and chiral materials using the Drude-Born-Fedorov model.  We find that the time-averaged Poynting vector always points along the wave vector,  the time-averaged energy-flux density is always positive, and the time-averaged energy density is positive (negative) when the refractive index is positive (negative).  The phase velocity is negative when the real part of the refractive index is negative, and the group velocity generally changes sign several times as a function of frequency near resonance.
\end{abstract}

\maketitle

\section{Introduction} \label{sec:Intro}
Negative refraction (NR) is a phenomenon in which electromagnetic waves are refracted at an interface with NR angle \cite{Veselago_68, Pendry_04, Smith_04, Pendry_06, Veselago_06}.  It is believed that in order for NR to occur, the real part of the (electric) permittivity ($\veps$) and real part of the (magnetic) permeability ($\mu$) must both be negative at a particular frequency \cite{Veselago_68, Pendry_04, Smith_04, Pendry_06, Veselago_06, Krowne_06, Eleftheriades_05, Ramakrishna_09, Capolino_09}.  Such materials are sometimes called ``double-negative'' materials.  NR meta-materials, i.e., specially designed NR materials made from assemblies of multiple elements fashioned from composite materials have been developed \cite{Smith_04, Krowne_06, Eleftheriades_05, Ramakrishna_09, Capolino_09}. It is furthermore believed that light at frequencies such that \{${\mathrm{Re}}[\veps(\omega)] > 0$ and ${\mathrm{Re}}[\mu(\omega)] < 0$\} or \{${\mathrm{Re}}[\veps(\omega)] < 0$ and ${\mathrm{Re}}[\mu(\omega)] > 0$\}, is not able to propagate in materials \cite{Veselago_68, Pendry_04, Smith_04, Pendry_06, Veselago_06, Krowne_06, Eleftheriades_05, Ramakrishna_09, Capolino_09}.  Both beliefs are predicated on the assumption that the permittivity and the permeability are real. However, this ``double-negative'' criterion does not apply for frequencies close to resonance, where the permittivity and the permeability are complex.  If the permittivity and the permeability are real and have different signs, the average energy flux density (average Poynting vector) vanishes (see Sec.~\ref{subsec:Poynting}).  However, if the permittivity and the permeability are complex, the average energy flux density is non-zero (the same is true for the chiral case treated in Sec.~\ref{chiral_admittance}).  Here we use the amplitude-phase representation of the permittivity and permeability within the Drude-Lorentz model for achiral media \cite{Drude-Lorentz} and the Drude-Born-Fedorov model \cite{BF} for chiral media to calculate the complex refractive index.  We then analyze and categorize the wealth of phenomena that occur when optical waves are in frequency ranges near resonant optical transitions where NR is possible.  The analysis presented here breaks the old paradigm for NR and creates a new one in which the frequency regime for NR is significantly broader when compared with previously deduced frequency ranges.

References \cite{Ziolkowski_01, McCall_02} are exceptions; they used the amplitude-phase representation of the permittivity, permeability and refractive index.  They consider two resultant complex refractive indices, $n(\omega) = \pm \sqrt{\veps(\omega) \mu(\omega)/(\veps_0 \mu_0)}$.  While the resultant complex refractive index $n(\omega)$ with the minus sign is not a physical solution, because it gives negative absorption (i.e., gain), the refractive index $n(\omega)$ with the plus sign has the right properties.

To date, NR materials have been fabricated using man-made meta-materials, but Ref.~\cite{Chen_21} argued that naturally occurring NR materials exist; the NR of n(?) were calculated.  NR meta-materials have led to significant technological advancements \cite{Pendry_06, Veselago_06, Krowne_06, Eleftheriades_05, Ramakrishna_09, Capolino_09} including: (1) superlensing, i.e., overcoming the diffraction limit of conventional lenses, allowing for sub-wavelength imaging for high-resolution microscopy \cite{Pendry_00, Zhang_08}, (2) cloaking using devices that can manipulate the flow of light around an object, rendering it invisible to observers \cite{Pendry_06,Ulf_10, Kim_15}, (3) terahertz imaging, spectroscopy, and communication systems, enabling non-invasive inspections in biomedical imaging and security screening \cite{Ling_18}, and (4) antennas incorporating NR meta-materials that can enhance the radiated power of the antenna NR by focusing electromagnetic radiation by a flat lens versus dispersion \cite{Elef_04, Ziol_03, Rennings_05}.

\section{Theory} \label{sec:Theory}
For an electromagnetic plane wave, with electric field $\mathbf{E}({\bf r},t) = \mathrm{Re} ( \boldsymbol{\mathcal{E}}_0 \, e^{i (\mathbf{k} \cdot \mathbf{r} - \omega t)} )$ and magnetic field $\mathbf{H}({\bf r},t) = \mathrm{Re} ( \boldsymbol{\mathcal{H}}_0 \, e^{i (\mathbf{k} \cdot \mathbf{r} - \omega t)} )$, the Faraday and Amp\`{e}re equations, together with the constitutive equations  in an isotropic homogeneous material, ${\bf D} = \veps {\bf E}$ and ${\bf B} = \mu {\bf H}$ yield, in SI units,
\begin{equation}  \label{eq:FaradayAmpere}
    {\bf k} \times \boldsymbol{\mathcal{E}}_0 = \omega \, \mu(\omega) \boldsymbol{\mathcal{H}}_0 ,
\quad
    {\bf k} \times \boldsymbol{\mathcal{H}}_0 = - \omega \, \veps(\omega) \boldsymbol{\mathcal{E}}_0 .
\end{equation}
Substituting ${\bf k} = \tfrac{n(\omega) \, \omega}{c} \, {\hat {\bf k}}$ we obtain
\begin{equation}  \label{eq:Faraday1}
    n(\omega) \, {\hat {\bf k}} \times \boldsymbol{\mathcal{E}}_0 = c \, \mu(\omega) \boldsymbol{\mathcal{H}}_0 ,
\end{equation}
\begin{equation}  \label{eq:Ampere1}
    n(\omega) \, {\hat {\bf k}} \times \boldsymbol{\mathcal{H}}_0 = - c \, \veps(\omega) \boldsymbol{\mathcal{E}}_0 ,
\end{equation}
which yields [noting that in vacuum, $c^2 = (\veps_0\mu_0)^{-1}$],
\begin{equation}  \label{eq:refindex}
n^2(\omega) = c^2 \veps(\omega) \mu(\omega) = \tfrac{\veps(\omega)\mu(\omega)}{\veps_0\mu_0}.
\end{equation}

\subsection{Drude-Lorentz model} \label{subsec:Drude-Lorentz}
The Drude-Lorentz model \cite{Drude-Lorentz} is a widely used theoretical framework for describing the behavior of electromagnetic waves in materials. It provides a phenomenological approach to model $\veps$ and $\mu$ of materials, including those with NR.  In the Drude-Lorentz model, the equation of motion for an electron in a meta-atom can be expressed as:
\begin{equation}  \label{eq:drude}
m \frac{d^{2}\mathbf{r}}{dt^{2}} = -m\omega_0^{2}\mathbf{r} - m\gamma \frac{d\mathbf{r}}{dt}+ (-e)\mathbf{E}(\omega)e^{-i\omega t},
\end{equation}
where $m$ is the effective mass of the electron, $\mathbf{r}(t)$ is the displacement of the electron, $\omega_0$ is the resonance frequency, $\gamma$ is the damping coefficient, $e$ is the elementary charge, and $\mathbf{E}(\omega)$ is the electric field of the incident electromagnetic wave at frequency $\omega$.  Substituting $\mathbf{r}(t) = \mathbf{r}_0(\omega) e^{-i \omega t}$, into Eq.~(\ref{eq:drude}), we find $\mathbf{r}_0(\omega) = \left(\frac{ -e/m}{\omega_0^{2} - \omega^{2} - i \gamma \omega}\right) \mathbf{E}(\omega)$.  The polarization $\mathbf{P}(\omega)$ related to the induced dipole moment per unit volume, can be written as $\mathbf{P}(\omega) = - N e \mathbf{r}_0(\omega) \equiv \chi(\omega) \mathbf{E}(\omega)$, where $\chi(\omega)$ is the electric susceptibility of the material.  Substituting the expression for $\mathbf{r}_0$ into $\mathbf{P}(\omega)$, we can obtain the electric susceptibility as $\chi(\omega) = -\frac{Ne^2}{m(\omega^2 - \omega_0^2 + i \gamma \omega)}$ where $N$ is the number density of electric dipole moments.   The electric permittivity $\veps(\omega)$ of the material can then be calculated as
\begin{equation}    \label{eq:veps}
\veps(\omega) = \veps_0(1 + \chi(\omega)) = \veps_0(1 - \frac{\omega_p^2}{\omega^2 - \omega_0^2 + i \gamma \omega}),
\end{equation}
where the plasma frequency is defined as $\omega_p^2 = \frac{Ne^2}{\veps_0 m}$ \cite{QM}.  [If we make use of the Clausius-Mossotti relation \cite{Bottcher-Lakhtakia}, then $\veps(\omega) = \veps_0(1 - \frac{\omega_p^2}{\omega^2 - \tilde \omega_0^2 + i \gamma \omega})$ where $\tilde \omega_0^2 = \omega_0^2 - \omega_p^2/3$, but the correction to the resonance frequency is usually small.] Similarly, a magnetic dipole transition with resonance frequency $\omega_{0m}$ and width $\gamma_m$ yields the magnetic permeability,
\begin{equation}    \label{eq:mu}
\mu(\omega) = \mu_0(1 - \frac{\omega_{pm}^2}{\omega^2 - \omega_{0m}^2 + i \gamma_m \omega}) ,
\end{equation}
where the magnetic plasma frequency squared $\omega_{pm}^2$ is a constant related to the magnetic properties of the material and is proportional to the transition magnetic dipole moment squared.

\subsection{Negative refractive index} \label{subsec:NRI}
In order to develop the theory of NR, Veselago \cite{Veselago_68} wrote, $n(\omega) = \pm \sqrt{\tfrac{\veps(\omega)\mu(\omega)}{\veps_0\mu_0}}$, where the minus sign is required for the case when the real parts of both $\veps(\omega)$ and $\mu(\omega)$ are negative.  This is the standard approach for dealing with double-negative materials \cite{Veselago_68, Pendry_04, Smith_04, Pendry_06, Veselago_06, Krowne_06, Eleftheriades_05, Ramakrishna_09, Capolino_09}.  Instead we follow a more direct and mathematically appealing procedure.  We write the complex refractive index as
\begin{equation}  \label{eq:n}
n(\omega) = \frac{\sqrt{|\veps(\omega)| |\mu(\omega)|}}{\sqrt{\veps_0\mu_0}} e^{i(\theta_\veps(\omega) + \theta_\mu(\omega))/2},
\end{equation}
where $\theta_\veps$ and $\theta_\mu$ are the complex phases of $\veps$ and $\mu$, respectively, i.e., $\veps = |\veps| e^{i \theta_\veps}$ and $\mu = |\mu| e^{i \theta_\mu}$. The real (imaginary) part of $n(\omega)$ is the refractive index (optical absorption coefficient divided by $\omega/c$).  The imaginary part of $n(\omega)$, $n''(\omega)$, must be non-negative since otherwise the plane wave for the electric field, $\mathbf{E}({\bf r},t) = \mathrm{Re} ( \boldsymbol{\mathcal{E}}_0 \, e^{i (\mathbf{k} \cdot \mathbf{r} - \omega t)} ) = \mathrm{Re} ( \boldsymbol{\mathcal{E}}_0 \, e^{i \omega [(n'(\omega)+i n''(\omega)) {\hat{\mathbf{k}}} \cdot \mathbf{r}/c - t]} )$, would have gain, which is impossible.  This form of the refractive index is unique (there are no branch point problems because of the square root in the definition of the refractive index, and because the imaginary part of $n(\omega)$ cannot be negative since that would imply gain).  Therefore, adding an angle $\pi$ to $\theta_n = (\theta_\varepsilon + \theta_\mu)/2$ in Eq.~(\ref{eq:n}) would yield an unphysical complex refractive index.

\subsection{Poynting vector} \label{subsec:Poynting}
The Poynting vector (which gives the electromagnetic energy transfer per unit area per unit time) is defined as $\mathbf{S} = \mathbf{E} \times \mathbf{H}$.  For a linearly polarized plane wave with $\mathbf{E}({\bf r},t) = \mathrm{Re} ( \boldsymbol{\mathcal{E}}_0 \, e^{i (\mathbf{k} \cdot \mathbf{r} - \omega t)} )$ and $\mathbf{H}({\bf r},t) = \mathrm{Re} ( \boldsymbol{\mathcal{H}}_0 \, e^{i (\mathbf{k} \cdot \mathbf{r} - \omega t)}  )$, where $\boldsymbol{\mathcal{E}}_0$ and $\boldsymbol{\mathcal{H}}_0$ obey the Eqs.~(\ref{eq:Faraday1}) and (\ref{eq:Ampere1}) with $\varepsilon$, $\mu$ and $n$ being complex, the three orthogonal vectors \{$\boldsymbol{\mathcal{E}}_0, \boldsymbol{\mathcal{H}}_0, \mathbf{k}$\} can be written as $\boldsymbol{\mathcal{E}}_0 = \! \mathcal{E}_0 \, \hat{\mathbf{x}}$, $\boldsymbol{\mathcal{H}}_0 = \mathcal{H}_0 \, \hat{\mathbf{y}}$ and $\mathbf{k} = k \, \hat{\mathbf{z}}$, where $k = \omega n(\omega) / c$, $\mathcal{H}_0 = \sqrt{ |\varepsilon| / |\mu| } \, \mathcal{E}_0 e^{i \theta_H}$, and $\theta_H = (\theta_{\veps} - \theta_{\mu})/2$.  The orthogonal vectors \{$\boldsymbol{\mathcal{E}}_0, \boldsymbol{\mathcal{H}}_0, \mathbf{k}$\} form a right- (left)-handed coordinate system if $\mathrm{Re}(n) >0$ ($\mathrm{Re}(n) < 0$). Taking $\mathcal{E}_0$ to be real and positive, we obtain
\begin{equation}  \label{eq:Poynting}
\mathbf{S} = \hat{\mathbf{z}} \, \sqrt{|\varepsilon| / |\mu| } \, \mathcal{E}_0^{2} \cos(\zeta) \cos(\zeta + \theta_H) e^{-2 k'' z} .
\end{equation}
Here $\zeta = k' z - \omega t$, and
$k' = n'(\omega)\omega/c$ ($k'' = n''(\omega)\omega/c$) is the real (imaginary) part of $k$, and $n'$ ($n''$) is the real (imaginary) part of $n$.
When $\theta_H$ vanishes, $\mathbf{S}$ is proportional to $\cos^2(\zeta)$ and is directed along 
$\hat{\mathbf{z}}$.  When $\theta_H \neq 0$, the magnetic field has phase shift $\theta_H$
with respect to the electric field.  Hence, there are intervals of $\zeta$ where both 
$\cos ( \zeta )$ and $\cos(\zeta + \theta_H)$ have the same sign and $\mathbf{S}$ is along $\hat{\mathbf{z}}$, 
and there are intervals of $\zeta$ where $\cos ( \zeta )$ and $\cos(\zeta + \theta_H)$ have different signs.  
In the latter case, $\mathbf{S}$ is along $-\hat{\bf z}$.
Hence the energy transfer per unit area per unit time is  time-dependent and can be either positive 
or negative. The time-averaged energy flux density is 
\begin{equation}  \label{eq:av_Poynting}
\bar { \mathbf{S} } \equiv \tfrac{1}{T} \int_{0}^{T} \mathbf{S}(z, t) d t  = \hat{\mathbf{z}} \, \sqrt{|\varepsilon| / |\mu| } \, 
(\mathcal{E}_0^{2}/2) \cos(\theta_H) e^{-2 k'' z} ,
\end{equation}
where $T = 2 \pi / \omega$ is the wave period.  If $\varepsilon$ and $\mu$ are real and have different signs, satisfying the Faraday and Amp\`{e}re equations requires $|\theta_H| = \pi/2$ and $\bar { \mathbf{S} } = 0$.  If $\varepsilon$ and $\mu$ are complex, $0 < \theta_{\varepsilon} < \pi$ and $0 < \theta_{\mu} < \pi$, and $|\theta_H| < \pi/2$, which is clear from the Drude-Lorentz model; $\bar{\mathbf{S}}(\omega)$ is always directed along $\hat{\mathbf{z}}$.

\subsection{Electromagnetic energy density and power dissipation density} \label{subsec:energy_density}

The time rate of change of the electromagnetic energy density is given by \cite{Jackson_99}
\begin{equation}   \label{eq:dU}
  \frac{\partial u}{\partial t} =
  \mathbf{E} \cdot \frac{\partial \mathbf{D}}{\partial t} +
  \mathbf{H} \cdot \frac{\partial \mathbf{B}}{\partial t} .
\end{equation}
Moreover, the rate of change of the energy density can be written as $\frac{\partial u}{\partial t} = Q + \frac{\partial u_{\mathrm{eff}}}{\partial t}$, where $Q$ is the power dissipation density, and $\frac{\partial u_{\mathrm{eff}}}{\partial t}$ is the energy density rate due to energy transfer,
i.e., the difference between incoming and outgoing electromagnetic energies.  For a narrow wave packet, the time-averaged power dissipation $\bar{Q}$ and the energy density $\bar{u}_{\mathrm{eff}}$ averaged over a period $2 \pi/\omega$ are \cite{Jackson_99,Webb-negative-energy-OptExpress-2012}
\begin{eqnarray}
  \bar{Q} &=&
  \frac{\omega}{2} \,
  \big[
    \veps'' (\omega) \,
    \big| \boldsymbol{\mathcal{E}}_{0} \big|^{2} +
    \mu'' (\omega) \,
    \big| \boldsymbol{\mathcal{H}}_{0} \big|^{2}
  \big]  e^{-2 k'' z} \nonumber \\
  &=&
  \frac{\omega}{2} \, \big\{
    \veps'' (\omega) + \mu'' (\omega) \,
    \big| \frac{\veps(\omega)}{\mu(\omega)} \big| \big\} \big| \boldsymbol{\mathcal{E}}_{0} \big|^{2}  e^{-2 k'' z} ,
  \label{eq:energy-dissipation}
  \\
  \bar{u}_{\mathrm{eff}} &=&
  \frac{1}{2} \,
  \bigg\{
    \frac{d (\omega \, \veps' (\omega))}{d \omega} \,
    \big| \boldsymbol{\mathcal{E}}_{0} \big|^{2}
    + \frac{d (\omega \, \mu' (\omega))}{d \omega} \,
    \big| \boldsymbol{\mathcal{H}}_{0} \big|^{2} \bigg\} e^{-2 k'' z} \nonumber \\
    &=&
  \frac{1}{2} \,
  \bigg\{
    \frac{d (\omega \, \veps' (\omega))}{d \omega} 
    + \frac{d (\omega \, \mu' (\omega))}{d \omega} \, \big| \frac{\veps(\omega)}{\mu(\omega)} \big| 
    \bigg\} \big| \boldsymbol{\mathcal{E}}_{0} \big|^{2} e^{-2 k'' z} . \nonumber \\
  \label{eq:energy-density}
\end{eqnarray}
Note that if $\omega_p > \sqrt{\gamma (2 \omega_0 + \gamma)}$, then $\veps' (\omega_{\pm}) = 0$, where
\begin{eqnarray*}
  \omega_{\pm} =
  \sqrt{\omega_{0}^{2} +
    \frac{\omega_{p}^{2} - \gamma^2}{2} \pm
    \sqrt{ \bigg( \frac{\omega_{p}^{2} - \gamma^2}{2} \bigg)^{2} - \omega_{0}^{2} \gamma^2}} .
\end{eqnarray*}
For $\omega_{-} < \omega < \omega_{+}$, $\veps' (\omega)$ is negative.
If $\omega$ is close to $\omega_{-}$, $\veps' (\omega)$ is small and $\frac{\partial \veps' (\omega)}{\partial \omega} < 0$,
then $\frac{d (\omega \, \veps' (\omega))}{d \omega} < 0$,
and the electric energy density $\frac{1}{2} \, \frac{d (\omega \, \veps' (\omega))}{d \omega} \, %
\big| \boldsymbol{\mathcal{E}}_{0} \big|^{2}$ is negative.
Moreover, if $\omega_{pm} > \sqrt{\gamma_m \, (2 \omega_{0m} + \gamma_m)}$, there is a range of $\omega$
where $\frac{d (\omega \, \mu' (\omega))}{d \omega} < 0$, and the magnetic energy density
$\frac{1}{2} \, \frac{d (\omega \, \mu' (\omega))}{d \omega} \, \big| \boldsymbol{\mathcal{H}}_{0} \big|^{2}$ is negative.

\subsection{Phase velocity and group velocity} \label{subsec:phase_group_velocity}
The phase velocity of light is $v_p = \omega / k'$, and the group velocity is
$v_g =  ( \partial k' / \partial \omega )^{-1}$.
Hence, $v_p = c / n'$, and $v_g = c / (n' + \omega \partial_{\omega} n')$.
Note that when $n' < 0$, the phase velocity is negative, and
when $n' + \omega \partial_{\omega} n' < 0$, the group velocity
is negative.  In other words, if a  pulse propagates through a material with a negative group velocity,
the peak of the pulse propagates in the direction opposite to the energy flow direction
\cite{Dolling-superluminal-Science-2006, Gehring-superluminal-Science-2006}.
Moreover, near a resonance, $n' + \omega \partial_{\omega} n'$ can be small,
and the group velocity can exceed the speed of light \cite{Segev-PhysRevA-2000}.
Experiments have verified that it is possible for the group velocity
to exceed the speed of light in vacuum \cite{Boyd-superluminal-Science-2009, Dolling-superluminal-Science-2006, Bigelow-JPhysics-CM-2006, Schweinsberg-superluminal-EPL-2006}.
From our calculations we see that the group velocity can become singular at as many as four frequencies.

A circular-polarization representation for achiral materials will be discussed following the chiral case below.  It turns out to be simpler than the linearly polarized analysis above.

\section{Numerical results for the achiral case} \label{sec:Numerical}
\begin{figure}%[ht]
\centering
\includegraphics[width=0.95\linewidth]{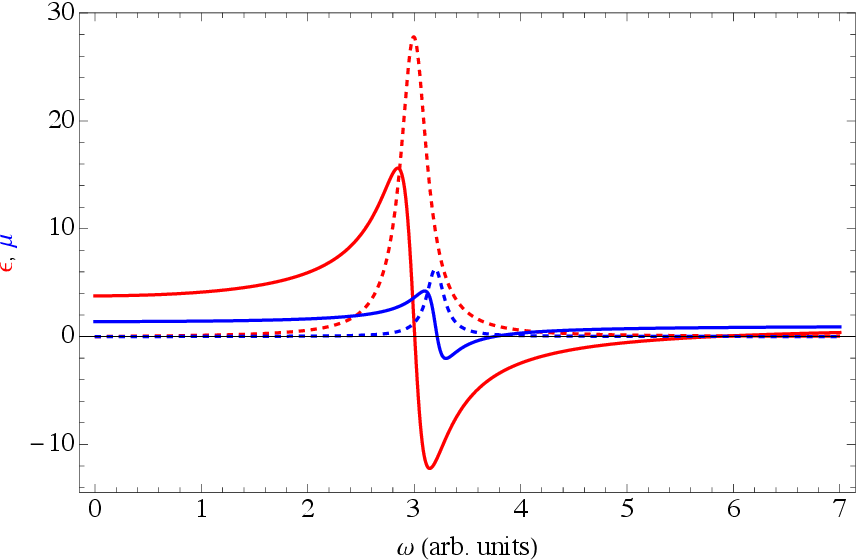}
\caption{The complex relative electric permittivity ($\veps/\veps_0$)  and complex magnetic permeability ($\mu/\mu_0$) plotted versus $\omega$ (in arbitrary units) for Drude-Lorentz parameters, $\omega_0 = 3, \omega_{0m} = 3.2, \gamma = 0.3, \gamma_m = 0.2, \omega_p = 5, and \omega_{pm} = 2$.  The dashed curves are the imaginary parts of $\veps(\omega)$ and $\mu(\omega)$.}
\label{Fig_permittivity_permeability}
\end{figure}

\begin{figure}%[htb]
\centering
\includegraphics[width=0.95\linewidth]{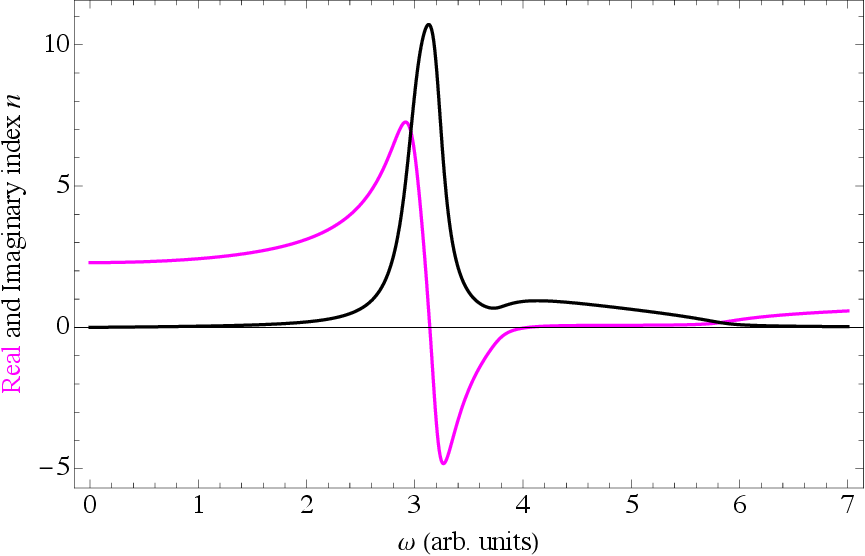}
\caption{The real (magenta) and imaginary (black) parts of the complex refractive index $n(\omega)$ versus $\omega$  (in arbitrary units) for the same Drude-Lorentz parameters as used in Fig.~\ref{Fig_permittivity_permeability}.  The NR region is to the right of the resonance frequency.  The imaginary part shows absorption over a large region about the resonance frequency.}
\label{Fig_refraction}
\end{figure}

Using the Drude-Lorentz model for an electric dipole transition and a magnetic dipole transition with resonance frequencies that are close to one another (see the caption of Fig.~\ref{Fig_permittivity_permeability} for the set of parameters used), we calculate the complex permittivity $\veps(\omega)$ and complex permeability $\mu(\omega)$ versus frequency, and the complex refractive index $n(\omega)$ (whose real part is the index of refraction and imaginary part is the absorption coefficient) using Eq.~(\ref{eq:n}).  Figure \ref{Fig_permittivity_permeability} shows the complex $\veps(\omega)$ and $\mu(\omega)$ versus frequency calculated with the Drude-Lorentz model.  The region of frequencies where both of the real parts of $\veps(\omega)$ and $\mu(\omega)$ are negative is shown in Fig.~\ref{Fig_permittivity_permeability}, but there are also frequency regions where the real parts of $\veps(\omega)$ and $\mu(\omega)$ are of opposite sign. 
Figure~\ref{Fig_refraction} plots the real and imaginary parts $n(\omega)$, defined in Eq.~(\ref{eq:n}), versus $\omega$.  Note that the absorption coefficient $\alpha \equiv \mathrm{Im}[n(\omega)] \ge 0$ for all frequencies [hence there is absorption (no gain) for all frequencies].  Moreover, there are frequency regions where $\mathrm{Re}[n(\omega)] < 0$ and in part of this frequency range the real parts of $\veps(\omega)$ and $\mu(\omega)$ are of {\it opposite} sign.

Note that all the figures in this paper appear in  the Supplemental Material (SM) \cite{SM}, which is a Mathematica notebook, that allows the reader to modify all the material parameters using the sliders.

\begin{figure}%[htb]
\centering
\includegraphics[width=0.9\linewidth]{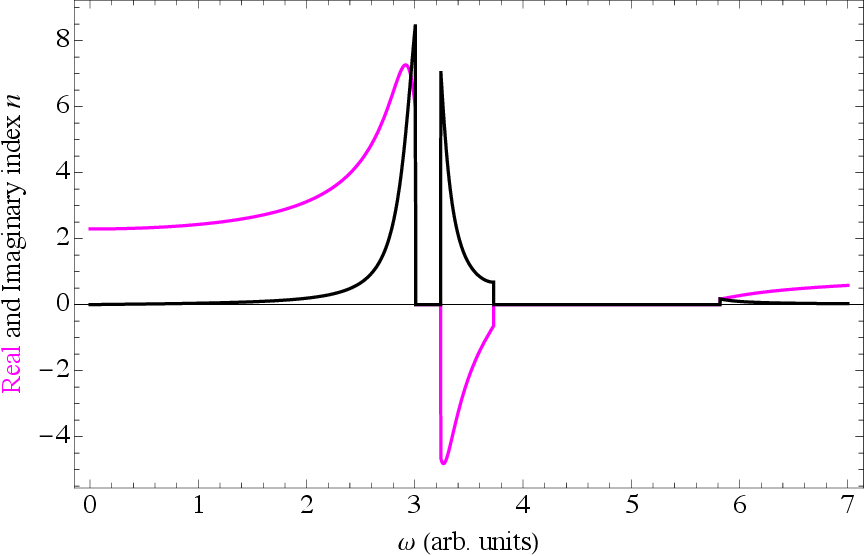}
\caption{The real and imaginary parts of the complex refractive index versus $\omega$ (in arbitrary units) for the same Drude-Lorentz parameters as used in Fig.~\ref{Fig_permittivity_permeability} using the ``double-negative'' criterion that propagation {\it cannot} occur if the real parts of $\veps$ and $\mu$ have different signs.  In the these frequency regions the real and imaginary parts of $n(\omega)$ have been set to zero, which is necessary according to the commonly held view.}
\label{Fig_refraction_Veselago}
\end{figure}

For comparison, Fig.~\ref{Fig_refraction_Veselago} shows $n(\omega)$ versus $\omega$ calculated assuming the light can propagate {\it only} if the real parts of $\veps$ and $\mu$ are of the same sign, as assumed in much of the literature.  Where they are of opposite sign, $n(\omega)$ has been set to zero in Fig.~\ref{Fig_refraction_Veselago}.  With the set of parameters used, there are two frequency regions where $n(\omega)$ has been set to zero; the frequency regions $[3.01, 3.24]$ and $[3.73, 5.82]$.   Readers can compare Fig.~\ref{Fig_refraction_Veselago} with Fig.~\ref{Fig_refraction} to see the behavior of the correct refractive index in these regions.  The large region where $n(\omega)$ is zeroed to the right of $\omega = 3.7$ is where the real part of $\veps$ is negative but the real part of $\mu$ is positive, and the same is true in the small region near $\omega = 3.0$ (see Fig.~\ref{Fig_permittivity_permeability}).

\begin{figure}%[ht]
\centering
\includegraphics[width=0.95\linewidth]{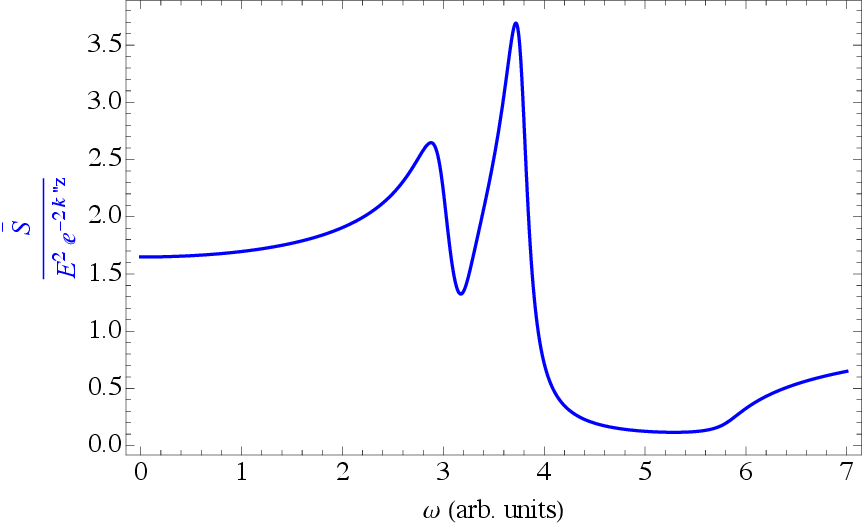}
\caption{The time-averaged Poynting vector $\bar{\mathbf{S}}$ plotted versus $\omega$ (in arbitrary units) for the same Drude-Lorentz parameters as used in Fig.~\ref{Fig_permittivity_permeability}.}
\label{Poynting-vector-achiral}
\end{figure}

\begin{figure}[ht]
\centering
\includegraphics[width=0.95\linewidth]{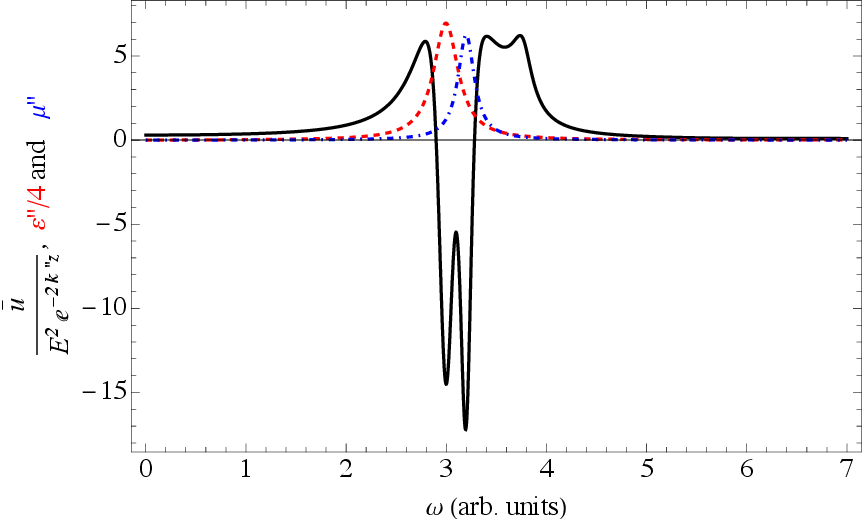}
\caption{The time-averaged energy density $\bar{u}(\omega)$ (black), the imaginary part of the permittivity, $\mathrm{Im} (\veps) \equiv \veps''$ (red)  and the imaginary part of the permeability, $\mathrm{Im} (\mu) \equiv \mu''$ (blue) versus $\omega$ (in arbitrary units) for the same Drude-Lorentz parameters as used in Fig.~\ref{Fig_permittivity_permeability}.}
\label{Fig_average-energy-density}
\end{figure}

Figure~\ref{Poynting-vector-achiral} plots the frequency dependence of the time-averaged Poynting vector, 
$\tfrac{\bar{\mathbf{S}}(\omega)}{\mathcal{E}_0^{2}e^{-2 k'' z}}$ [which is equivalent to the non-averaged 
quantity $\tfrac{\mathbf{S}_{\pm}(\omega)}{\mathcal{E}_{\pm}^2 e^{-2 k''_{\sigma} z}}$
for right and left polarized light in the {\it chiral case}, as discussed in Sec.~\ref{sec:Chiral}].
The average Poynting vector is positive (i.e., in the $z$ direction) for all frequencies.  The peaks of ${\bar S}$ appear
at the frequencies of the maximum and minimum of $n(\omega)$, respectively.  Figure~\ref{Fig_average-energy-density} plots both the average energy density $\bar u(\omega)$ and the imaginary parts of the permittivity and permeability, $\veps''(\omega) \equiv \mathrm{Im}[\veps (\omega)]$ and $\mu''(\omega) \equiv \mathrm{Im}[\mu (\omega)]$, versus frequency.  $\bar u(\omega)$ is negative in the frequency region near the maximum of $\veps''$, and in a region near the maximum of $\mu''$.

\begin{figure}%[ht]
\centering
\includegraphics[width=0.95\linewidth]{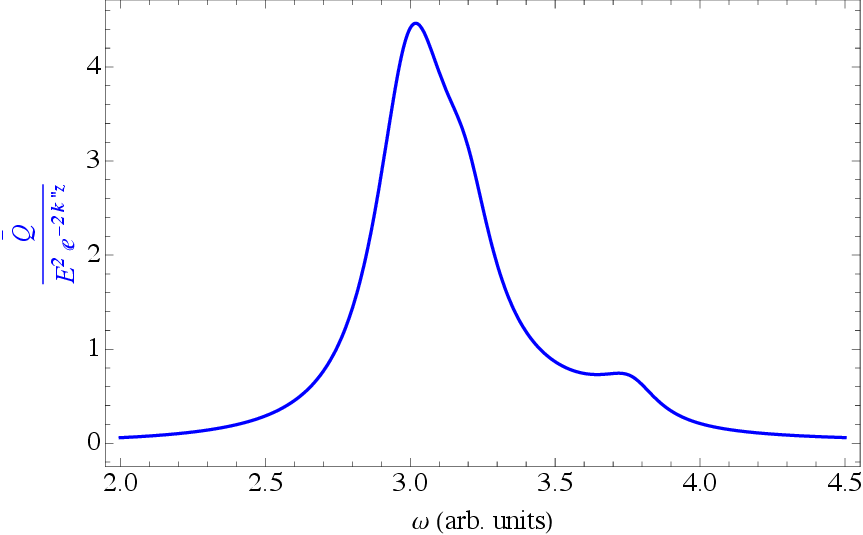}
\caption{The time-averaged power dissipation density $\bar{Q}(\omega)$ versus $\omega$ (in arbitrary units) for the same Drude-Lorentz parameters as used in Fig.~\ref{Fig_permittivity_permeability}.}
\label{Fig_average-power-dissipation-density}
\end{figure}

Figure~\ref{Fig_average-power-dissipation-density} shows the time-averaged power dissipation density $\bar{Q}(\omega)$ versus $\omega$ given in Eq.~(\ref{eq:energy-dissipation}).  The major peak near $\omega = 3$ is due to the peak of the imaginary part of $n(\omega)$ and the minor peak near $\omega = 3.8$ is near the small peak of the imaginary part of $n(\omega)$ near $\omega = 3.9$.  In Fig.~\ref{Fig_permittivity_permeability} we can see that the imaginary part of $\mu(\omega)$ has a peak at $\omega = 3.2$ which contributes to the peak of  $\bar{Q}(\omega)$ but is shifted due to the factor of $\big| \frac{\veps(\omega)}{\mu(\omega)} \big|$ multiplying it [see Eq.~(\ref{eq:energy-dissipation})].

\begin{figure}[ht]
\centering
\includegraphics[width=0.95\linewidth]{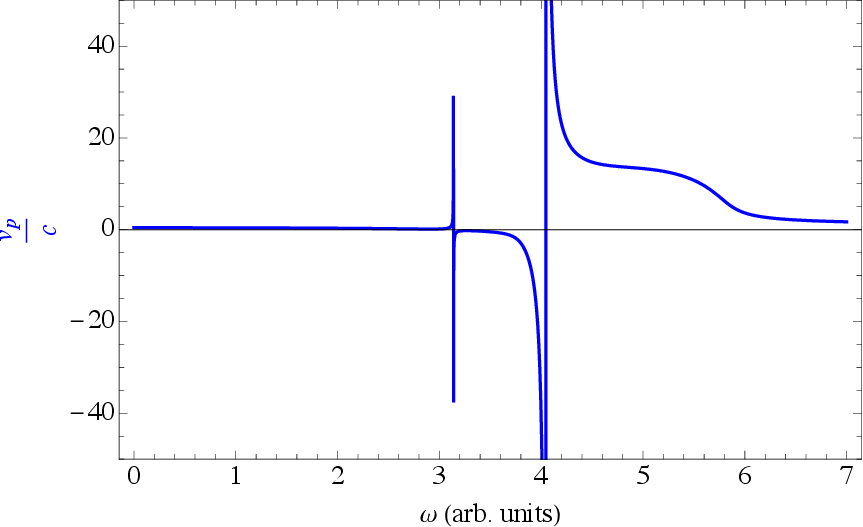}
\caption{Phase velocity $v_p = \omega / k'$ versus frequency, for the same Drude-Lorentz parameters as used in Fig.~\ref{Fig_permittivity_permeability}.}
\label{phase_velocity}
\end{figure}

\begin{figure}[ht]
\centering
\includegraphics[width=0.95\linewidth]{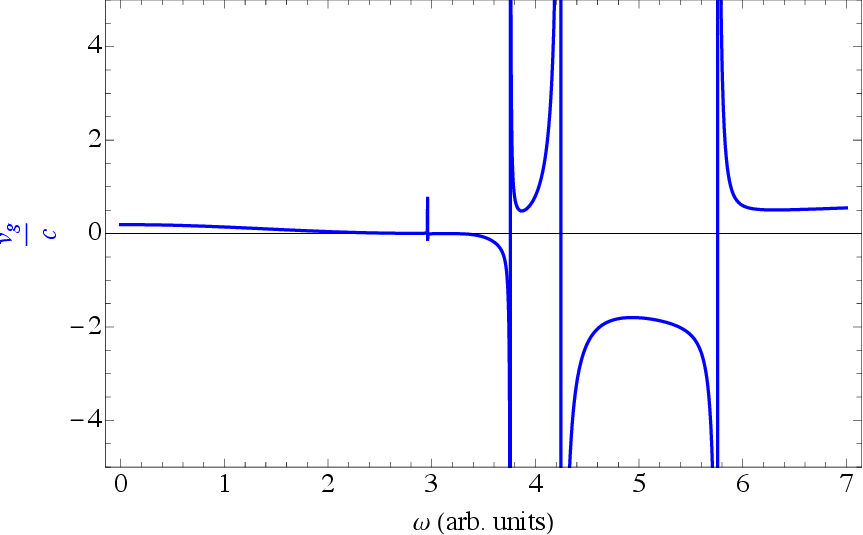}
\caption{Group velocity $v_g =  ( \partial k' / \partial \omega )^{-1}$ versus frequency, for the same Drude-Lorentz parameters as used in Fig.~\ref{Fig_permittivity_permeability}.}
\label{group_velocity}
\end{figure}

Figure \ref{phase_velocity} plots the phase velocity $v_p = \omega / k'$ versus frequency.  There are two resonances in the phase velocity that originate from the two zeros in $\mathrm{Re}[n(\omega)] \equiv n'(\omega)$.  Figure \ref{group_velocity} plots the group velocity $v_g =  ( \partial k' / \partial \omega )^{-1}$ versus frequency.   There are four resonances for the group velocity that originate from four zeros in $\partial [\omega n'(\omega)] / \partial \omega$.

\begin{figure}[ht]
\centering
\includegraphics[width=0.95\linewidth]{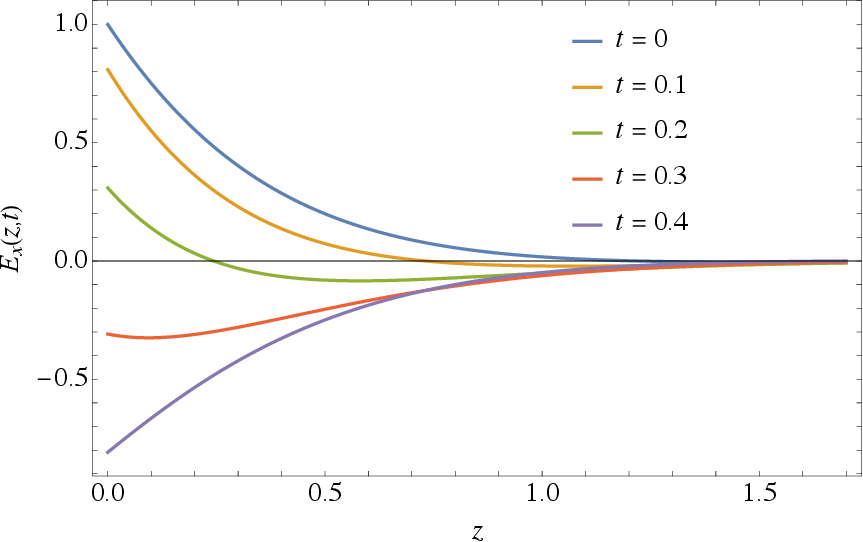}
\caption{Simulation of the electric field $E_x(z,t) = \mathrm{Re} [ \mathcal{E}_{x} (z, t) ]$ of the plane wave propagating along the $z$-axis plotted versus $z$ at five times, $t = 0, 0.1, 0.2, 0.3, 0.4$ (arb. units), where the period $\tau = 2 \pi/\omega = 1$.  Here we took $\omega = 3.8$ (arb. units) and the same Drude-Lorentz parameters as used in Fig.~\ref{Fig_permittivity_permeability}, so $n(\omega) = -0.339 + 0.731 i$.}
\label{simulation_Ex}
\end{figure}

Figure \ref{simulation_Ex} simulates a plane wave which is polarized along the $x$-axis as it propagates along the $z$-axis.  The figure shows the electric field $\mathbf{E}({\bf r},t) = {\hat {\bf x}} \, \mathrm{Re} [e^{i (k z - \omega t)}]$, where $k = n(\omega) \omega/c$, versus $z$ at five different times.  At a frequency $\omega = 3.8$ (arb. units), the phase of the electric field propagates in the $-z$ direction (the curves shift left with increasing time), hence the real part of the refractive index is negative (see below). For the Drude-Lorentz parameters used here, the permittivity is $\veps (\omega) = -3.402 + 0.923 \, i$, and the permeability is $\mu (\omega) = 0.078 + 0.167 \, i$, i.e., $\veps' (\omega)$ and $\mu' (\omega)$ have different signs (so according to the ``double-negative'' criterion that propagation {\it cannot} occur if the real parts of $\veps$ and $\mu$ have different sign, propagation cannot occur at this frequency), and the refractive index is $n(\omega) = -0.339 + 0.731 i$.  In the section titled ``Simulation of the propagation of a plane wave in an achiral medium'' of the SM \cite{SM} you can run a movie of the propagation of the electric field and watch it propagate by opening the time slider and playing the movie.  In the SM \cite{SM} you can also change the frequency $\omega$ as well as other Drude-Lorentz parameters to see how the propagation changes as the parameters are varied.

\section{Chiral Media} \label{sec:Chiral}

The constitutive equations for isotropic chiral media must be modified to allow for optical activity. One form of the modified constitutive equations, called the Drude-Born-Fedorov model \cite{Drude-Lorentz, BF}, is as~follows:
\begin{eqnarray}
\boldsymbol{\mathcal{D}} &=& \veps[\boldsymbol{\mathcal{E}} + \beta \, \nabla \times \boldsymbol{\mathcal{E}}] , \label{eqn:2.172}\\
\boldsymbol{\mathcal{B}} &=& \mu [\boldsymbol{\mathcal{H}} + \beta \, \nabla \times \boldsymbol{\mathcal{H}}] , \label{eqn:2.173}
\end{eqnarray}
where the electric permittivity $\veps (\omega)$ is given in Eq.~(\ref{eq:veps}), and the magnetic permeability $\mu (\omega)$ is given in Eq.~(\ref{eq:mu}).  This form is symmetric under time-reversal. The pseudoscalar $\beta$, sometimes called the chiral admittance, has the units of length and is a measure of the optical activity. Let us consider a plane wave electromagnetic field, ${\bf E}({\bf r},t) = {\mathrm{Re}}[\boldsymbol{\mathcal{E}}({\bf r},t)] = {\mathrm{Re}}[\boldsymbol{\mathcal{E}}_0 e^{i ({\bf k} \cdot {\bf r} - \omega \, t)}]$, and similarly for ${\bf B}({\bf r},t)$, ${\bf D}({\bf r},t)$ and ${\bf H}({\bf r},t)$, and determine the consequences of Eqs.~(\ref{eqn:2.172})-(\ref{eqn:2.173}). Using the Faraday and Amp\`{e}re laws in a nonconducting medium,
$\nabla \times \boldsymbol{\mathcal{E}} = i\omega \boldsymbol{\mathcal{B}}$, $\nabla \times \boldsymbol{\mathcal{H}} = -i\omega \boldsymbol{\mathcal{D}}$,
the Drude-Born-Fedorov equation in frequency space takes the form
\begin{eqnarray}    \label{eqn:2.175}
\boldsymbol{\mathcal{D}}_0 = \veps(\omega) \left[\boldsymbol{\mathcal{E}}_0 + i\omega \beta(\omega) \boldsymbol{\mathcal{B}}_0\right], \\ \nonumber
 \boldsymbol{\mathcal{B}}_0 = \mu(\omega) \left[\boldsymbol{\mathcal{H}}_0 - i\omega \beta(\omega)\boldsymbol{\mathcal{D}}_0\right] .
\end{eqnarray}
These equations can be written in matrix form as
\begin{eqnarray}
&&\left(\begin{matrix}\boldsymbol{\mathcal{E}}_0\\
\boldsymbol{\mathcal{H}}_0\\
\end{matrix}\right)  = 
\left(\begin{matrix}
\veps^{-1}(\omega) &- i\omega \beta(\omega) \\
i\omega \beta(\omega) &\mu^{-1}(\omega) 
\end{matrix}\right) 
\left(\begin{matrix}\boldsymbol{\mathcal{D}}_0\\
\boldsymbol{\mathcal{B}}_0
\end{matrix}\right). \, \, \, \, \, \, \, \, \label{eqn:DBF}
\end{eqnarray}
Substituting into the Faraday and Amp\`{e}re equations gives
\begin{equation}
{\bf k} \times \boldsymbol{\mathcal{E}}_0 = \omega \boldsymbol{\mathcal{B}}_0, \qquad 
{\bf k} \times \boldsymbol{\mathcal{H}}_0 = -\omega \boldsymbol{\mathcal{D}}_0,
\end{equation}
which, upon writing ${\bf k} = \tfrac{\tilde n(\omega) \omega}{c} \, {\hat {\bf k}}$, can be written in terms of the refractive index as
\begin{eqnarray}
\frac{\tilde n(\omega)}{c} \, {\hat {\bf k}} \times (\veps^{-1}\boldsymbol{\mathcal{D}}_0 - i\omega \beta \boldsymbol{\mathcal{B}}_0) &=& 
\boldsymbol{\mathcal{B}}_0,  \nonumber  \\
\frac{\tilde n(\omega)}{c} \, {\hat {\bf k}} \times (\mu^{-1} \boldsymbol{\mathcal{B}}_0 + i\omega \beta \boldsymbol{\mathcal{D}}_0) &=& - 
\boldsymbol{\mathcal{D}}_0.
\end{eqnarray}
Solving for $\tilde n(\omega)$ in the determinant obtained using these equations yields two degenerate solutions for the right-and left-handed circularly polarized light fields,
\begin{equation}  \label{eq:nchiral}
 \tilde n(\omega) \equiv \ n_\pm(\omega) = \frac{n(\omega)}{1 \mp \frac{\beta \omega}{c} \, n(\omega)} ,
\end{equation}
where $n(\omega)$ is given in Eq.~(\ref{eq:n}).

\subsection{Non-resonant chiral admittance}  \label{chiral_admittance}

In this subsection we consider a chiral medium containing impurity atoms with an electric dipole transition and impurity atoms with a magnetic dipole transition.  The chiral admittance $\beta$ of the chiral medium has resonances that are very far from the resonance frequencies of the atoms, so it is reasonable to take $\beta$ as frequency-independent \cite{Varadan_94}.  In the next subsection we investigate the case where the chiral admittance has a resonance form \cite{Condon_37, Wiltshire_09, Zhao_10, Oh_15}.

Using a circular polarization basis we write
\begin{equation}   \label{eq:circular-polarization}
  \boldsymbol{\mathcal{E}}_0 =
  \mathcal{E}_{+} \, \mathbf{e}_{+} +
  \mathcal{E}_{-} \, \mathbf{e}_{-} ,
\end{equation}
(similarly for $\boldsymbol{\mathcal{D}}_0$, $\boldsymbol{\mathcal{B}}_0$ and $\boldsymbol{\mathcal{H}}_0$)
where the subscript $\pm$ refers to right-polarized (left-polarized) waves,
$\mathbf{e}_{+} = \frac{-1}{\sqrt{2}} \, \big( \hat{\mathbf{x}} + i \hat{\mathbf{y}} \big)$, and
$\mathbf{e}_{-} = \frac{1}{\sqrt{2}} \, \big( \hat{\mathbf{x}} - i \hat{\mathbf{y}} \big)$.
We find that
\begin{eqnarray}
  &&
  \mathcal{D}_{\pm} = \frac{n_{\pm}(\omega)}{n(\omega)} \, \varepsilon(\omega) \mathcal{E}_{\pm} ,
  \quad
  \mathcal{B}_{\pm} = \mp \frac{i n_{\pm}(\omega)}{c} \, \mathcal{E}_{\pm} ,
  \nonumber \\ &&
  \mathcal{H}_{\pm} =
  \mp i \sqrt{ \frac{|\varepsilon(\omega)|}{| \mu(\omega) |} } \,
  e^{i \theta_H}
  \mathcal{E}_{\pm} .
  \label{eq:B-right-vs-D}
\end{eqnarray}
The real part of the complex wavenumber of a circularly polarized wave is $k'_{\sigma} = \omega n'_{\sigma}(\omega) / c$, where $\sigma = \pm$, and $n'_{\sigma}(\omega)$ is the real part of $n_{\sigma}(\omega)$.  The rotation angle of the polarization of linearly polarized light is given by $\theta_{\mathrm{rot}} \equiv {\mathrm{Re}}[\Delta n] l = (n'_{+} - n'_{-}) l$, where $l$ is the length of the NR material traversed.  Moreover, differential absorption $\Delta \alpha_\pm \equiv {\mathrm{Im}}[\Delta n] \omega/c = (n''_{+} - n''_{-}) \omega/c$ (i.e.,the circular dichroism) ensues, hence the light will generally be elliptically polarized upon propagation through the material, and  $\theta_{\mathrm{rot}}(\omega)$ will be the rotation of the major and minor axes.

\begin{figure}%[htb]
\centering
\includegraphics[width=0.95\linewidth]{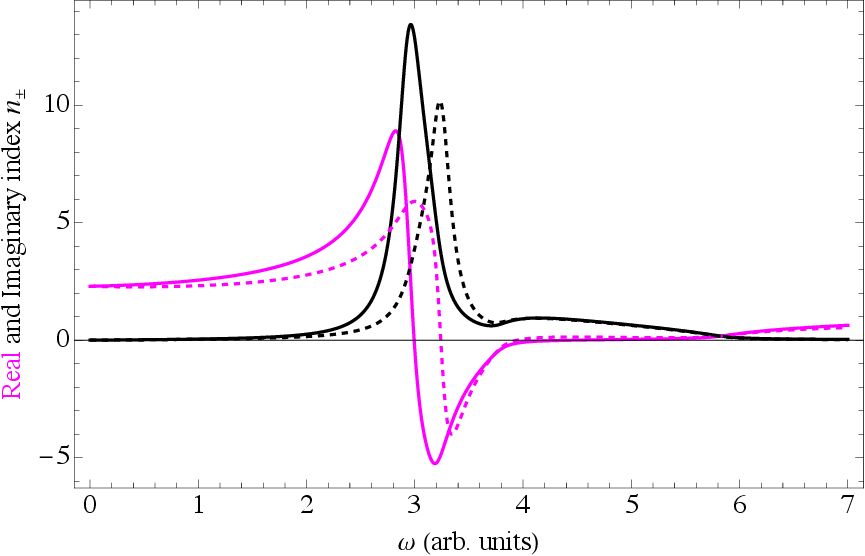}
\caption{In a medium that is chiral, the real and imaginary parts of the complex refractive index $n_\pm(\omega)$ versus $\omega$ (in arbitrary units) are different for right-circular polarization (solid curves) and left (dashed). The same Drude parameters as used in Fig.~\ref{Fig_permittivity_permeability} are used here, and $\beta = 0.02$ (arb. units).}
\label{Fig_refraction_chiral}
\end{figure}

Figure~\ref{Fig_refraction_chiral} plots the real (magenta) and imaginary (black) parts of $n_\pm(\omega)$ versus $\omega$ for right-circular polarized ($+$) (solid curves) and left-circular polarized ($-$) light (dashed curves).  The resonance frequency and the entire curves for the complex refractive index are shifted to higher (lower) frequencies for the right-circular polarized (left-circular polarized) light in the chiral medium (finite $\beta$) relative to those in Fig.~\ref{Fig_refraction} (which is for $\beta = 0$).  Moreover, both ${\mathrm{Re}}(n_+)$ and ${\mathrm{Im}}(n_+)$ are smaller in magnitude than ${\mathrm{Re}}(n_-)$ and ${\mathrm{Im}}(n_-)$, respectively.  Both ${\mathrm{Re}}(n_+)$ and ${\mathrm{Re}}(n_-)$ have regions of NR, and these regions partly overlap.  Both ${\mathrm{Im}}(n_\pm)$ are positive (absorptive) for all frequencies.  The absorption of both right and left circular polarized light have minima near $\omega \approx 3.8$.

\begin{figure}%[htb]
\centering
\includegraphics[width=0.95\linewidth]{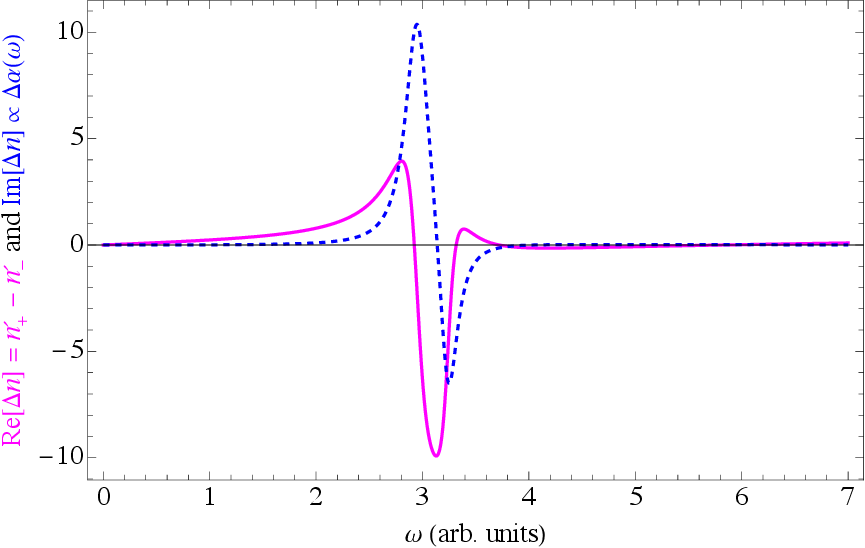}
\caption{In a chiral medium, the real and imaginary parts of the complex refractive index $n_\pm(\omega)$ versus $\omega$ (in arbitrary units) are different for right-circular polarization (solid curves) and left (dashed). The same Drude-Born-Fedorov parameters are used here as in Fig.~\ref{Fig_refraction_chiral}.}
\label{Fig_Birefringence_optical_activity}
\end{figure}

 Figure \ref{Fig_Birefringence_optical_activity} plots the circular birefringence ${\mathrm{Re}}[\Delta n(\omega)] = n'_{+}(\omega) - n'_{-}(\omega)$, which is promotional to the polarization rotation angle $\theta_{\mathrm{rot}}(\omega) = [n'_{+}{\mathrm{rot}}(\omega)  - n'_{-}{\mathrm{rot}}(\omega)] l$  versus frequency, and the circular dichroism ${\mathrm{Im}}[\Delta n(\omega)] = n''_{+}(\omega) - n''_{-}(\omega)$, which is proportional to the differential absorption $\Delta \alpha(\omega) \equiv (n''_{+}(\omega) - n''_{-}(\omega)) \omega/c$ versus frequency.  Both the circular birefringence and the circular dichroism change sign as a function of frequency near the resonance, but the real part changes sign twice whereas the imaginary part changes sign only once.  For frequencies larger than $\omega \approx 3.8$, both the circular birefringence and the circular dichroism are very small (the solid and dashed curves in Fig.~\ref{Fig_refraction_chiral} are very close to overlapping).

The possibility of obtaining negative refraction in a chiral medium due to a resonance in the permittivity {\it only} (and not in the permeability) was first suggested by Pendry \cite{Pendry_Sci_04}; see also Refs.~\cite{Monzon_05, Tretyakov_05}.  This can be understood from Eq.~(\ref{eq:nchiral}) by noting that if the real part of the second term in the denominator is larger than unity, $n_+$ would be negative (while $n_-$ would be positive).  In Fig.~\ref{Fig_refraction_chiral} we see that there are frequency ranges where both $n_+$ and $n_-$ are negative due to the resonance structure of $\veps$ and $\mu$.  The resonance in $n_+$ occurs at smaller frequency than the resonance in $n_-$.

The Poynting vector is given by
\begin{equation}   \label{eq:Poynting-right}
  \mathbf{S}_{\pm}(\omega) = 
  \hat{\mathbf{k}} \,
  \sqrt{ \frac{| \varepsilon(\omega) |}{| \mu(\omega) |} } \,
 \frac{ \mathcal{E}_{\pm}^2 }{2} \,
  \cos \theta_H e^{-2 k''_{\sigma} z} .
\end{equation}
Since $|\theta_H| \leq \pi/2$, $\mathbf{S}_{\pm}$ points in the $\hat{\mathbf{k}}$ direction.  Note that the energy flux density $\frac{\mathbf{S}_{\pm}}{e^{-2 k''_{\sigma} z}}$ does not depend on $\mathbf{r}$ or $t$ (as opposed to the non-time averaged achiral case).  The Poynting vector $\tfrac{\mathbf{S}_{\pm}(\omega)}{\mathcal{E}_{\pm}^2 e^{-2 k''_{\sigma} z}}$, is independent of the polarization $\sigma$ and the chiral admittance $\beta$, there is no need to present a figure for this case since the results are identical to Fig.~\ref{Poynting-vector-achiral} which shows the average energy flux density $\bar{\mathbf{S}}$ obtained in the achiral case discussed in Sec.~\ref{sec:Theory}.

Following a similar line of reasoning as used by Jackson \cite{Jackson_99} for the achiral case, we find that the energy density $u_{\pm}$ and the power dissipation density $Q_{\pm}$ are given by \cite{Poulikakos_16}
\begin{eqnarray}
  u_{\pm} &=&
  \frac{1}{16 \pi}
  \bigg\{
    \frac{d \big[ \omega \tilde\veps' (\omega)]}{d \omega} \,
    \big| \boldsymbol{\mathcal{E}}_{\pm} (\omega) \big|^{2} +
    \frac{d \big[ \omega \tilde\mu' (\omega)]}{d \omega} \,
    \big| \boldsymbol{\mathcal{H}}_{\pm} (\omega) \big|^{2}
  \nonumber \\ && +
    2 \frac{d \big[ \omega^2 \tilde\beta' (\omega)]}{d \omega} \,
    \mathrm{Im}
    \big[
      \boldsymbol{\mathcal{H}}_{\pm} (\omega)
      \cdot
      \boldsymbol{\mathcal{E}}_{\pm}^{*} (\omega)
    \big]
  \bigg\}
  \nonumber \\ &=&
  \frac{\big| \boldsymbol{\mathcal{E}}_{\pm} (\omega) \big|^{2}}{16 \pi}
  \Bigg\{
    \frac{d \big[ \omega \tilde\veps' (\omega)]}{d \omega} +
    \frac{d \big[ \omega \tilde\mu' (\omega)]}{d \omega} \,
    \Big| \frac{\veps (\omega)}{\mu (\omega)} \Big|
  \nonumber \\ && \mp
    2 \, \frac{d \big[ \omega^2 \tilde\beta' (\omega)]}{d \omega} \,
    \mathrm{Re}
    \bigg[
      \sqrt{\frac{\veps (\omega)}{\mu (\omega)}}
    \bigg]
  \Bigg\} , \\
  Q_{\pm} &=&
  \frac{\omega}{8 \pi} \,
  \Big\{
    \tilde\veps'' (\omega) \,
    \big| \boldsymbol{\mathcal{E}}_{\pm} (\omega) \big|^{2} +
    \tilde\mu'' (\omega) \,
    \big| \boldsymbol{\mathcal{H}}_{\pm} (\omega) \big|^{2}
    \nonumber \\ && +
    2 \omega \tilde\beta'' (\omega) \,
    \mathrm{Im}
    \big[
      \boldsymbol{\mathcal{H}}_{\pm} (\omega)
      \cdot
      \boldsymbol{\mathcal{E}}_{\pm}^{*} (\omega)
    \big]
  \Big\}
  \nonumber \\ &=&
  \frac{\omega \big| \boldsymbol{\mathcal{E}}_{\pm} (\omega) \big|^{2}}{8 \pi} \,
  \Bigg\{
    \tilde\veps'' (\omega) +
    \tilde\mu'' (\omega) \,
    \Big| \frac{\veps (\omega)}{\mu (\omega)} \Big|
    \nonumber \\ && \mp
    2 \, \omega \tilde\beta'' (\omega) \,
    \mathrm{Re}
    \bigg[
      \sqrt{\frac{\veps (\omega)}{\mu (\omega)}}
    \bigg]
  \Bigg\} .
\end{eqnarray}
Note that circularly polarized light has a constant magnitude, hence time-averaging the energy density and power dissipation density is not necessary when a circular polarization basis is used (unlike the linearly polarization basis used in the achiral section).

\begin{figure}%[htb]
\centering
\includegraphics[width=0.95\linewidth]{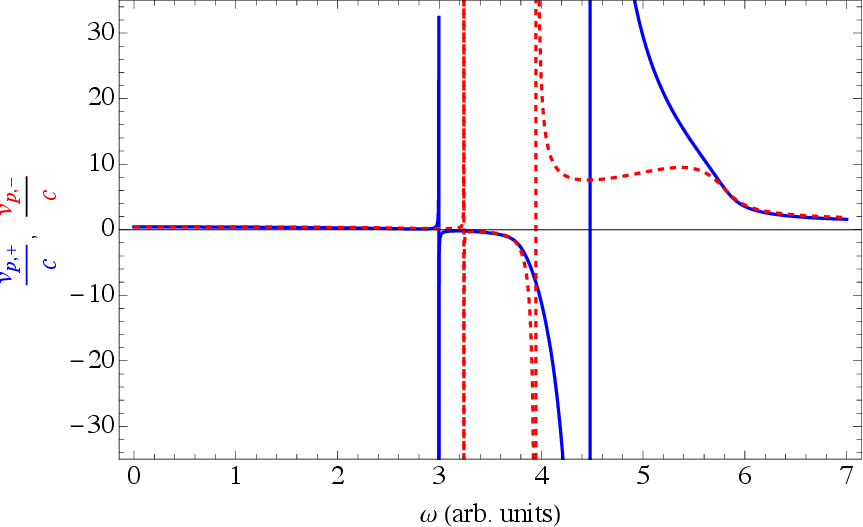}
\caption{In a chiral medium, the phase velocities $v_{p,\pm}(\omega)$ versus $\omega$ are different for right-circular polarization (solid curves) and left (dashed).  The same Drude-Born-Fedorov parameters are used here as in Fig.~\ref{Fig_refraction_chiral}.}
\label{Fig_phase-velocity-chiral}
\end{figure}

\begin{figure}%[htb]
\centering
\includegraphics[width=0.95\linewidth]{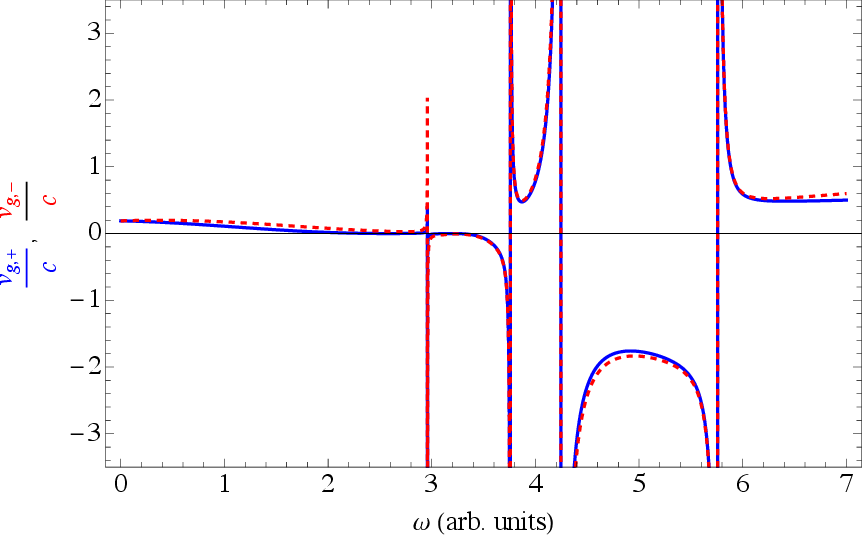}
\caption{In a medium that is chiral, the phase velocities $v_{g,\pm}(\omega)$ versus $\omega$ are different for right-circular polarization (solid curves) and left (dashed).  The same Drude-Born-Fedorov parameters are used here as in Fig.~\ref{Fig_refraction_chiral}.}
\label{Fig_group-velocity-chiral}
\end{figure}

In the limit $\beta \to 0$, the phase velocities $v_{p, \sigma}(\omega)$ and group velocities $v_{p, \sigma}(\omega)$ go to the achiral ones, and, as already stated, the energy flux density and the energy density go to the achiral {\it average} energy flux density and 
the {\it average} energy density, respectively.

The phase velocity is $v_{p, \sigma}(\omega) = \omega/ k'_{\sigma} = \frac{c}{n'_{\sigma}(\omega)}$, and the group velocity is $v_{g, \sigma} (\omega) = ( \partial k'_{\sigma} / \partial  \omega )^{-1} = \frac{c}{n'_{\sigma}(\omega) + \omega \partial_{\omega} n'_{\sigma}(\omega)}$.  Figure \ref{Fig_phase-velocity-chiral} plots the phase velocities $v_{p,\pm}(\omega)$ versus $\omega$ and Fig.~\ref{Fig_group-velocity-chiral} plots the group velocities $v_{g,\pm}(\omega)$.  The resonance behavior of the phase and group velocities are similar to the achiral case; the phase velocities have two resonances and the group velocities have four.  The SM notebook \cite{SM} allows you vary the $\beta$ parameter (and other parameters) to see the dramatic changes in the frequency dependence  of $\theta_{\mathrm{rot}}(\omega)$, $\Delta \alpha_\pm$ and other quantities such as the phase and group velocities.

\begin{figure}[htb]
\centering
\includegraphics[width=0.95\linewidth]{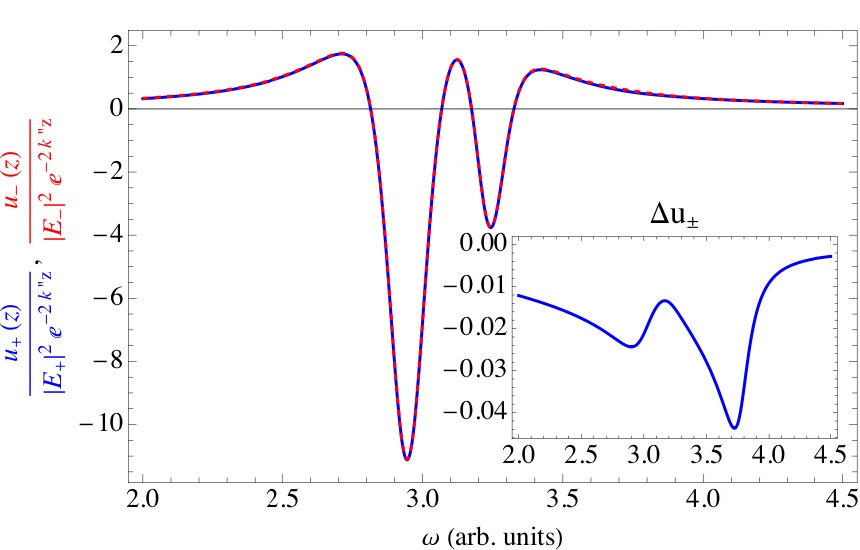}
\caption{In a chiral medium, the energy density $u_{\pm}(\omega)$ versus $\omega$ for right-circular (red) and left-circular polarization (dashed blue).  The same Drude-Born-Fedorov parameters are used here as in Fig.~\ref{Fig_refraction_chiral}.  The inset shows the difference $\Delta u_{\pm}(\omega) = \tfrac{u_{+}(\omega)(\omega)}{\mathcal{E}_{+}^{2} e^{-2 k''_{\sigma} z}} - \tfrac{u_{-}(\omega)(\omega)}{\mathcal{E}_{-}^{2} e^{-2 k''_{\sigma} z}}$.}
\label{Fig_energy-density-chiral}
\end{figure}

\begin{figure}[htb]
\centering
\includegraphics[width=0.95\linewidth]{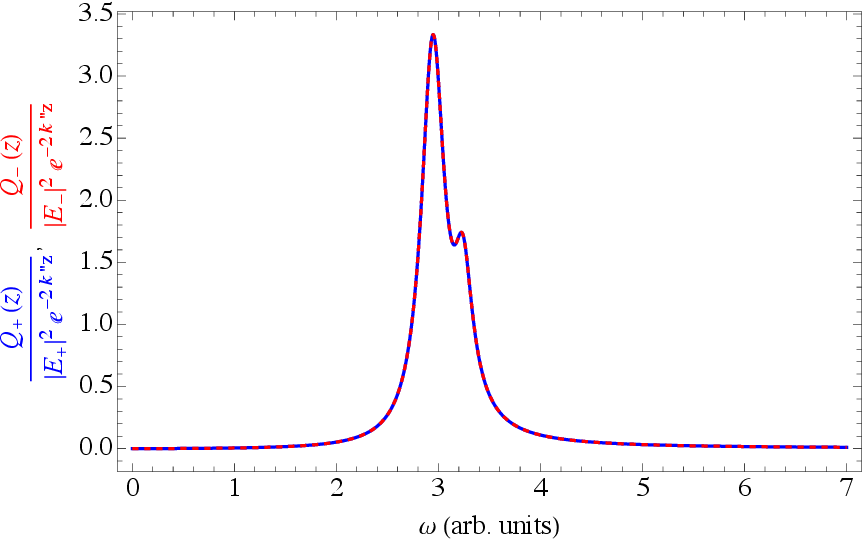}
\caption{In a chiral medium, the power dissipation density $Q_{\pm}(\omega)$ versus $\omega$ for right-circular (red) and left-circular polarization (dashed blue).  The difference $\Delta Q_{\pm}(\omega) = \tfrac{Q_{+}(\omega)(\omega)}{\mathcal{E}_{+}^{2} e^{-2 k''_{\sigma} z}} - \tfrac{Q_{-}(\omega)(\omega)}{\mathcal{E}_{-}^{2} e^{-2 k''_{\sigma} z}}$ is identically zero.  The same Drude-Born-Fedorov parameters are used here as in Fig.~\ref{Fig_refraction_chiral}.}
\label{Fig_power-dissipation-density-chiral}
\end{figure}

Figure \ref{Fig_energy-density-chiral} plots $u_{+}(\omega)$ and $u_{-}(\omega)$ versus frequency.  The right and left polarization curves are hard to distinguish but the inset plots the difference $\Delta u_{\pm}(\omega) = \tfrac{u_{+}(\omega)(\omega)}{\mathcal{E}_{+}^{2} e^{-2 k''_{+} z}} - \tfrac{u_{-}(\omega)(\omega)}{\mathcal{E}_{-}^{2} e^{-2 k''_{-} z}}$.  Figure \ref{Fig_power-dissipation-density-chiral} shows the power dissipation density $Q_{\pm}(\omega)$ versus $\omega$ for right-circular (red) and left-circular polarization (dashed blue).  The power dissipation is always positive, and resonance behavior has two local maxima in this case, and the difference $\Delta Q_{\pm}(\omega) = \tfrac{Q_{+}(\omega)(\omega)}{\mathcal{E}_{+}^{2} e^{-2 k''_{+} z}} - \tfrac{Q_{-}(\omega)(\omega)}{\mathcal{E}_{-}^{2} e^{-2 k''_{-} z}} = 0$ for all frequencies if $\beta$ is real (as we shall see in Sec.~\ref{subsec:resonant_beta}, for complex $\beta$, $\Delta Q_{\pm}(\omega)$ is in general not zero).

\begin{figure}[htb]
\centering
\includegraphics[width=0.95\linewidth]{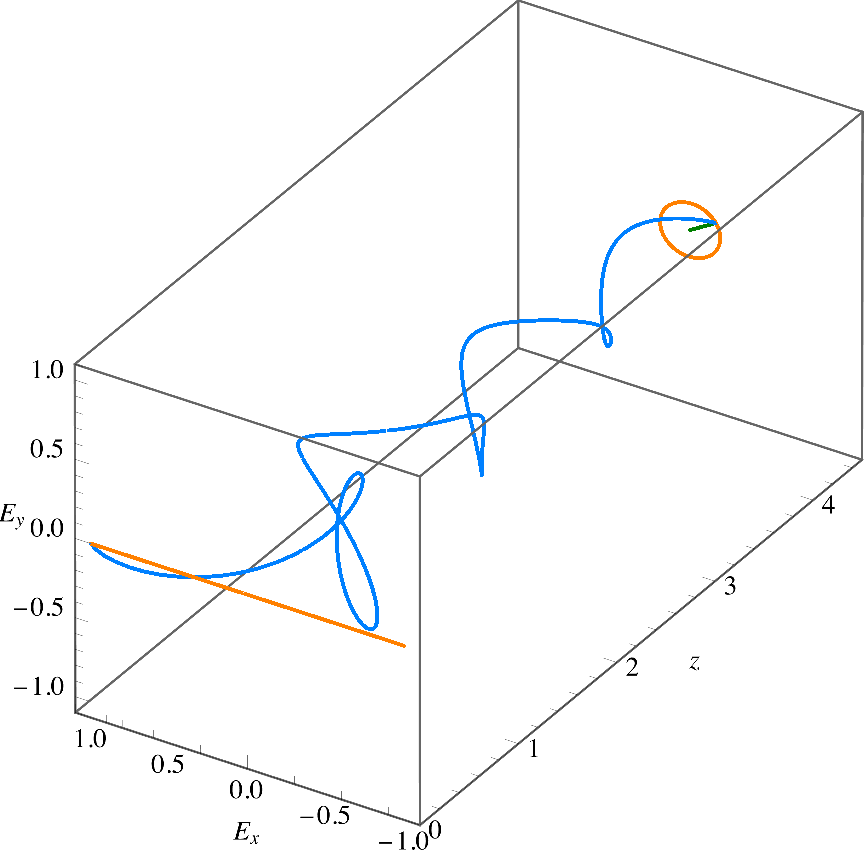}
\caption{The electric field ${\bf E}(z, t)$ of the plane wave (\ref{E-polarized}) propagating along the $z$-axis plotted as a function of $z$ for $t = 0$.  The orange line at the front face of the material shows the linear polarization of the incident light.  The intersection of the blue curve with the orange line gives the instantaneous polarization at $t = 0$.  The orange ellipse shows the polarization states of the light at the back face of the material as a function of time, and the intersection of the blue curve with the orange ellipse gives the instantaneous polarization at $t = 0$. The same Drude-Born-Fedorov parameters are used here as in Fig.~\ref{Fig_refraction_chiral}.  In the SM \cite{SM} you can run a movie showing the time-dependence of the polarization.}
\label{circ_dichroism}
\end{figure}

Finally, Fig.~\ref{circ_dichroism} shows the electric field ${\bf E}(z, t)$ of a plane wave propagating along the $z$-axis plotted versus $z$ (in the SM \cite{SM} you can run a movie which shows the time-dependence of ${\bf E}(z, t)$, and you can also change the material parameters). The light field at $z=0$ is linearly polarized (see orange line at the front face of the material in the figure).  The polarization becomes elliptical as the field propagates in $z$.  The electric field can be represented as
\begin{equation}   \label{E-polarized}
{\bf E}(z, t) = 2^{-1/2} \mathcal{E}_{0} {\mathrm{Re}}[(- \mathbf{e}_{+} e^{i k_+ z} + \mathbf{e}_{-} e^{i k_- z}) e^{-i\omega t}].
\end{equation}
In addition to the electric field inside the material, the linear polarization of the incident field at the front face of the medium is determined by the intersection of the blue curve with the orange line at the front face, and the polarization ellipse of the field at $z = z_{\mathrm{max}}$ is shown at the back face as the intersection of the blue curve with the green line and the orange ellipse.  To better understand the plane wave polarization dynamics, the reader is strongly encouraged to use the Manipulate command in the SM \cite{SM}.

\subsection{Resonant chiral admittance} \label{subsec:resonant_beta}

The chiral admittance $\beta (\omega)$ can be frequency dependent \cite{Condon_37, Wiltshire_09}; it can even have a resonance form (that can be modeled by a Lorentzian function) \cite{Zhao_10, Oh_15},
\begin{equation}   \label{eq:beta-Drude}
  \beta (\omega) = \frac{c \, \omega_{p \beta}}{\omega^2 - \omega_{0 \beta}^{2} + i \omega \gamma_{\beta}} .
\end{equation}
Here $\omega_{0 \beta}$ is the resonance frequency, $\gamma_{\beta}$ is the resonance width, and $\omega_{p \beta}$ is the chiral admittance strength parameter.  We use a model similar to Ref.~\cite{Condon_37} wherein we have a chiral medium which has an electric dipole transition with susceptibility $\chi_{\beta} (\omega)$ having resonance frequency $\omega_{0\beta}$ that results in a contribution to the permittivity with resonance frequency $\omega_{0\beta}$ and resonance width $\gamma_{\beta}$, and the chiral medium has impurity atoms. The impurity atoms have resonant electric susceptibility $\chi(\omega) = \veps(\omega)/\veps_0 -1$, where $\veps(\omega)$ is given in Eq.~(\ref{eq:veps}).  Furthermore, the chiral medium (without impurity atoms) has susceptibility
\begin{equation}   \label{eq:veps-correctuion}
  \chi_b (\omega) = - \frac{\omega_{pb}^{2}}{\omega^2 - \omega_{0\beta}^{2} + i \omega \gamma_{\beta}} ,
\end{equation}
where $\omega_{pb}$ is a plasma frequency of the medium without impurity atoms.  The total permittivity of the chiral medium with impurity atoms is
\begin{equation}   \label{eq:permittivity-chiral-resonant}
  \tilde\veps (\omega) = \veps_0 \,  [ 1 + \chi (\omega) + \chi_{\beta} (\omega) ] .
\end{equation}
The permeability $\mu(\omega)$ of impurity atoms is given in Eq.~(\ref{eq:mu}).  Note that if $\omega_{0\beta}$ is of the same order of magnitude as $\omega_0$ and $\omega_{0m}$, the optical characteristics of the resonant chiral materials can be very different than the constant $\beta$ case.  Specifically, if $\omega$ is close to $\omega_{0\beta}$, the chiral admittance can be large.  Appendix~\ref{appendix} discusses the conditions that restrict the parameters appearing in the resonance chiral admittance given in Eq.~(\ref{eq:beta-Drude}).

\begin{figure}%[htb]
\centering
\includegraphics[width=0.95\linewidth]{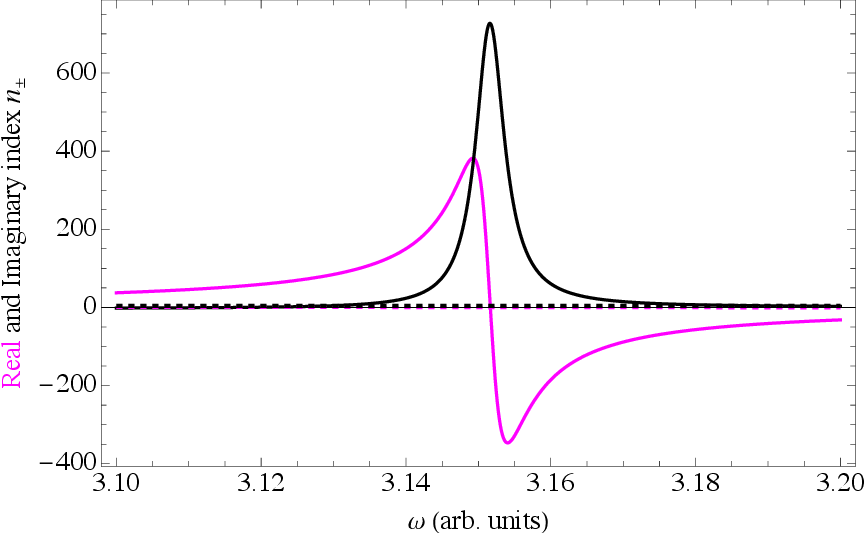}
\caption{Refractive index for a medium with resonant chiral admittance.  The real and imaginary parts of the complex refractive index $n_\pm(\omega)$ versus $\omega$ (in arbitrary units) are different for right-circular polarization (solid curves) and left (dashed). The same Drude parameters for $\veps$ and $\mu$ as used in Fig.~\ref{Fig_permittivity_permeability} are used here, and the parameter for $\beta(\omega)$ are $\omega_{0\beta} = 3.3$, $\omega_{p\beta} = 0.4$, $\gamma_{\beta} = 4.3$.}
\label{Fig_n+n-_resonance_chiral}
\end{figure}

\begin{figure}%[htb]
\centering
\includegraphics[width=0.95\linewidth]{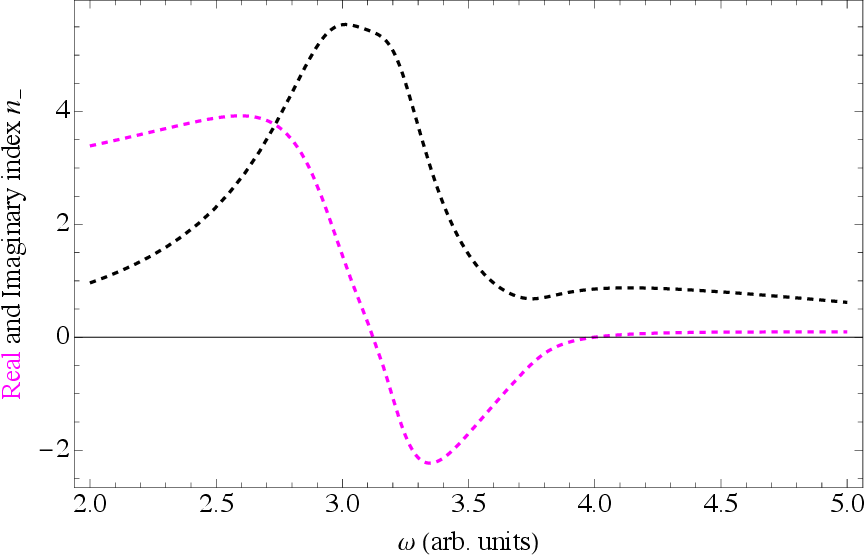}
\caption{Blowup of the ordinate in Fig.~\ref{Fig_n+n-_resonance_chiral} showing the real and imaginary parts of $n_-(\omega)$ versus frequency.}
\label{Fig_nminus_refraction_resonance_chiral}
\end{figure}

Figure \ref{Fig_n+n-_resonance_chiral} shows the real and imaginary parts of the refractive index in Eq.~(\ref{eq:nchiral}), with $\beta$ on the right hand side of the equation replaced by $\beta(\omega)$ given by Eq.~(\ref{eq:beta-Drude}) for a medium with a resonance in the chiral admittance very close to the resonance frequencies of the permittivity and permeability.  The right circularly polarized light has huge $n'_+(\omega)$ which changes sign at $\omega \approx 3.15$ and a gigantic $n''_+(\omega)$ (hence absorption) with a maximum at the same frequency.  For left circularly polarized light, both $n'_-(\omega)$ and $n''_-(\omega)$ cannot be seen on the ordinate scale of the figure, therefore we plot these quantities in Fig.~\ref{Fig_nminus_refraction_resonance_chiral}.  The refractive index $n'_-(\omega)$ is negative in the frequency region $3.12 < \omega < 3.98$, and the absorption has a maximum at $\omega \approx 3.03$.  Comparing the refractive index here with that shown in Fig.~\ref{Fig_refraction_chiral} for the non-resonance $\beta$ case, we see that the scale of the right-circularly polarized refractive index is roughly two orders of magnitude larger here, but the left-circularly polarized refractive index is the same order of magnitude here.

\begin{figure}[htb]
\centering
\includegraphics[width=0.95\linewidth]{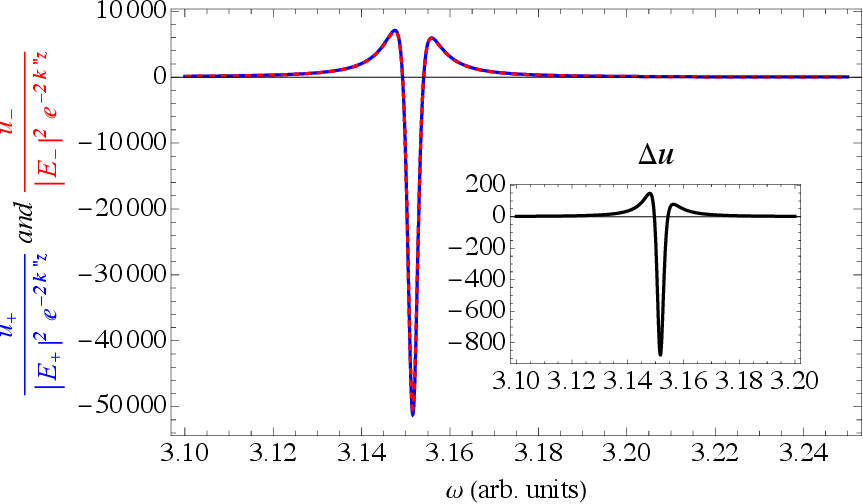}
\caption{Energy density $u_{\pm}(\omega)$ versus $\omega$ for right-circular (red) and left-circular polarization (dashed blue) for a chiral medium with resonant $\beta(\omega)$.  The inset shows a blowup of difference $\Delta u_{\pm}(\omega) = \tfrac{u_{+}(\omega)(\omega)}{\mathcal{E}_{+}^{2} e^{-2 k''_{\sigma} z}} - \tfrac{u_{-}(\omega)(\omega)}{\mathcal{E}_{-}^{2} e^{-2 k''_{\sigma} z}}$. The same Drude-Born-Fedorov parameters as used in Fig.~\ref{Fig_n+n-_resonance_chiral} are used here.}
\label{Fig_energy-density_resonance_chiral}
\end{figure}

Figure \ref{Fig_energy-density_resonance_chiral} shows the energy densities $u_{\pm}(\omega)$ versus $\omega$ for right-circularly and left-circularly polarized light.  There are two maxima and one minimum in the energy density.  In comparing with the energy density in Fig.~\ref{Fig_energy-density-chiral} we see that again there is a region where the energy density is negative, but the shape of the energy density curve as a function of frequency is very different here.  Because of the resonance in $\beta$, the minimum is three orders of magnitude deeper than in the non-resonance case.  The inset shows that $\Delta u(\omega) = \tfrac{u_{+}(\omega)(\omega)}{\mathcal{E}_{+}^{2} e^{-2 k''_{+} z}} - \tfrac{u_{-}(\omega)(\omega)}{\mathcal{E}_{-}^{2} e^{-2 k''_{-} z}}$ is four orders of magnitude larger than the non-resonant $\beta$ case. 

\begin{figure}[htb]
\centering
\includegraphics[width=0.95\linewidth]{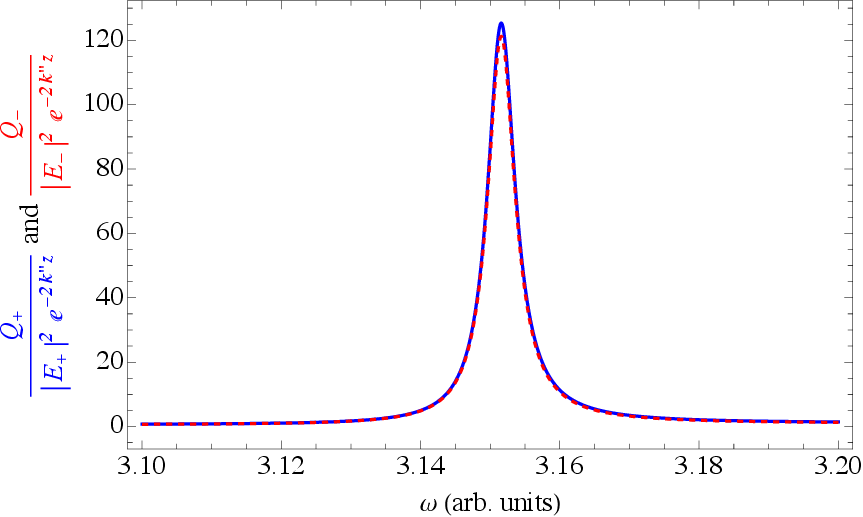}
\caption{In a chiral medium with resonant $\beta(\omega)$, the power dissipation densities $Q_{\pm}(\omega)$ versus $\omega$ for right-circular (red) and left-circular polarization (dashed blue).  The difference $\Delta Q_{\pm}(\omega) = \tfrac{Q_{+}(\omega)(\omega)}{\mathcal{E}_{+}^{2} e^{-2 k''_{\sigma} z}} - \tfrac{Q_{-}(\omega)(\omega)}{\mathcal{E}_{-}^{2} e^{-2 k''_{\sigma} z}}$ is very small and positive in this case; the difference is largest at $\omega \approx 3.152$ where $\Delta Q_{\pm} = 4.05$. The same Drude-Born-Fedorov parameters as used in Fig.~\ref{Fig_n+n-_resonance_chiral} are used here.}
\label{Fig_power-dissipation-density_resonance_chiral}
\end{figure}

Figure \ref{Fig_power-dissipation-density_resonance_chiral} shows the power dissipation densities $Q_{\pm}(\omega)$ versus $\omega$.  The maxima for $Q_{\pm}(\omega)$ occur at $\omega \approx 3.152$, but the $Q_{+}(\omega)$ maximum is a little larger and at a slightly higher frequency that for $Q_{-}(\omega)$.  The power dissipation densities here are roughly a factor of 40 larger than for the non-resonant $\beta$ case shown in Fig.~\ref{Fig_power-dissipation-density-chiral}.

\section{Summary} \label{sec:Summary}
We calculated the complex $\veps$, $\mu$ and refractive index $n$ versus frequency using a Drude-Lorentz model for a material having electric dipole and magnetic dipole transition resonances that are near one another, using the amplitude-phase representation for complex functions in Eq.~(\ref{eq:n}) to define $\sqrt{\veps \mu}$.  We then evaluated the phase velocity, group velocity, Poynting vector (energy flux density), energy density and power dissipation density, and discussed their surprising behavior for frequencies near the resonances.  We then treated chiral media using the Drude-Born-Fedorov model.  The circular polarized representation used to treat the chiral case determines the optical rotation activity and circular dichroism of the light given incident linearly polarized light, and yields the Poynting vector (energy flux density) and energy density which are independent of position and time (this is true for the time-averaged quantities in the achiral case too).  In the limit as the chiral admittance $\beta \to 0$, the energy flux density and energy density are identical to the temporally averaged energy flux density and energy density calculated with the linear polarized representation in the achiral case. We then considered the case when the chiral admittance $\beta$ depends on frequency and has a resonance near the resonance frequencies of the permittivity and permeability of the media, and saw that the results can vary dramatically from those obtained when the chiral admittance is almost constant near the resonance frequencies of the permittivity and permeability.  We intend to generalize the theory to treat non-isotropic condensed matter systems in a future publication.

\appendix
\section{Conditions on the parameters appearing in the resonant chiral admittance} \label{appendix}

Let us now consider a medium without any impurity atoms in order to determine the conditions the restrict the parameter values appearing in $\beta(\omega)$. 
The electric susceptibility of the medium is given in Eq.~(\ref{eq:veps-correctuion}), the medium permittivity is $\veps_{b}(\omega) = \veps_0 (1 + \chi_{b}(\omega))$, and the chiral admittance is given by Eq.~(\ref{eq:beta-Drude}).
The refractive index of the chiral medium without impurity atoms is
\begin{equation}   \label{eq:n-chiral-pure}
  n_{b \pm} (\omega) =
  \frac{n_b (\omega)}{1 \mp \frac{\beta (\omega) \, \omega}{c} \, n_b(\omega)} .
\end{equation}
where
\begin{equation}   \label{eq:n-medium-achiral}
  n_b (\omega) = \sqrt{\veps_b(\omega)} .
\end{equation}
The imaginary part of the refractive index must be positive, $n''_{\beta \pm} (\omega) \ge 0$,
which implies
\begin{equation}   \label{eq:n_Im>beta_Im}
  n''_b (\omega) > \omega \, \beta''(\omega) \, |n_b(\omega)|^{2} .
\end{equation}
Assuming that $\gamma_{\beta}$ is large, $\gamma_{\beta} \gg \omega_{o\beta}$, and applying the condition  (\ref{eq:n_Im>beta_Im}) to the range $\omega \ll \gamma{\beta}$  we get the inequality
\begin{equation}   \label{eq:beta-Im-condition-series}
  \omega_{p\beta} < \frac{\omega_{pb}^{2}}{2 \omega} .
\end{equation}
This inequality is satisfied in the frequency range $\omega < \omega_{\max} = \omega_{pb}^{2}/(2\omega_{p\beta})$.
On the other hand, when $\omega_{\max} \gg \omega_{0\beta}$,
$$
  \frac{\omega_{pb}^{2}}{2 \omega_{p\beta}} \gg \omega_{0\beta} .
$$

\end{document}